\documentclass[traditabstract]{aa} 
\usepackage{graphicx}
\usepackage{txfonts}
\begin{document}
   \title{Physical Properties of OSIRIS-REx Target Asteroid (101955) 1999~RQ$_{36}$}

   \subtitle{derived from Herschel\thanks{{\it Herschel}
      is an ESA space observatory with science instruments provided by
      European-led Principal Investigator consortia and with important
      participation from NASA.}, ESO-VISIR\thanks{Analysis is also based
      on observations collected at the European Southern Observatory, Chile;
      ESO, No.\ 287.C-5045(A)} and Spitzer observations}

   \author{T.\ G.\ M\"uller\inst{1} \and
           L.\ O'Rourke\inst{2} \and
           A.\ M.\ Barucci\inst{3} \and
           A.\ P\'al\inst{4} \and
           C.\ Kiss\inst{4} \and
           P.\ Zeidler\inst{5} \and
           B.\ Altieri\inst{2} \and
           B.\ M.\ Gonz\'alez-Garc{\'\i}a\inst{6} \and
           M.\ K\"uppers\inst{2}
          }

   \institute{Max-Planck-Institut f\"ur extraterrestrische Physik (MPE),          
              Giessenbachstrasse, 85748 Garching, Germany;
              \email{tmueller@mpe.mpg.de} \and
              European Space Astronomy Centre (ESAC), European Space Agency,                     
              Apartado de Correos 78, 28691 Villanueva de la Ca\~nada,
              Madrid, Spain \and
              Observatoire de Paris, Laboratoire d'Etudes Spatiales et            
              d'Instrumentation en Astrophysique (LESIA),
              5 Place Jules Janssen, 92195 Meudon Cedex, France \and
              Konkoly Observatory, Research Center for Astronomy and Earth Sciences, Hungarian Academy of Sciences; 
              Konkoly Thege 15-17, H-1121 Budapest, Hungary \and
              Ludwig-Maximilians-Universit\"at M\"unchen,                         
              Fakult\"at f\"ur Physik, Schellingstra{\ss}e 4,
              D-80799 M\"unchen, Germany \and
              INSA at European Space Astronomy Centre (ESAC), European Space Agency,                     
              Apartado de Correos 78, 28691 Villanueva de la Ca\~nada,
              Madrid, Spain
             }

   \date{Received ; accepted }

\abstract{In September 2011, the Herschel Space Observatory performed an
  observation campaign with the PACS photometer observing the asteroid
  \object{(101955) 1999~RQ$_{36}$} in the far infrared. The Herschel observations
  were analysed, together with ESO VLT-VISIR and Spitzer-IRS data,
  by means of a thermophysical model
  in order to derive the physical properties of 1999~RQ$_{36}$.
  We find the asteroid has an effective diameter in the range
  480 to 511\,m,
  a slightly elongated shape with a semi-major axis ratio of a/b=1.04,
  a geometric albedo of 0.045$^{+0.015}_{-0.012}$, and a retrograde rotation
  with a spin vector between -70 and -90$^{\circ}$ ecliptic latitude.
  The thermal emission at wavelengths below 12\,$\mu$m -originating in
  the hot sub-solar region- shows that there may be large variations
  in roughness on the surface along the equatorial zone of 1999~RQ$_{36}$, but
  further measurements are required for final proof.
  We determine that the asteroid has a disk-averaged thermal inertia of
  $\Gamma$=650\,Jm$^{-2}$s$^{-0.5}$K$^{-1}$ with a 3-$\sigma$ confidence range
  of 350 to 950\,Jm$^{-2}$s$^{-0.5}$K$^{-1}$, equivalent to what is observed
  for \object{25143~Itokawa} and suggestive that \object{1999~RQ$_{36}$}
  has a similar surface texture
  and may also be a rubble-pile in nature. The low albedo indicates that
  \object{1999~RQ$_{36}$} very likely contains primitive volatile-rich material,
  consistent with its spectral type, and that it is an ideal target for the
  OSIRIS-REx sample return mission.}


\authorrunning{M\"uller et al.}
\titlerunning{Thermal \& Shape properties of OSIRIS-REx Asteroid (101955) 1999~RQ$_{36}$}

   \maketitle
%

\section{Introduction}

 On May 25, 2011, NASA announced the selection of OSIRIS-REx as the third selected mission
 of its New Frontiers Program. The OSIRIS-REx mission will be
 launched in September 2016, fly by the Earth for a gravity assist
 in September 2017, and encounter the near-Earth asteroid (NEA)
 (101955) 1999~RQ$_{36}$ (Lauretta et al.\ \cite{lauretta10})
 in November 2019. Proximity operations at the
 asteroid last through March 2021, when it will acquire up to 2\,kg
 sample of its surface material. The sample return capsule will return
 to the Earth's surface in September 2023. 

 Based on its visible spectrum, \object{1999~RQ$_{36}$} is classified
 as a B-type and CM chondrite meteorites are considered to be the 
 corresponding spectral analog (Clark et al.\ \cite{clark11}).
 B-class asteroids are found mostly
 in the middle and outer regions of the main belt and are believed to be primitive
 and volatile-rich. However, the Polana family in the inner belt contains
 B-types, and dynamical studies have shown that 1999~RQ$_{36}$
 might be a liberated member of this family,
 considered as an important inner-belt source of low-albedo NEAs
 (Campins et al.\ \cite{campins10}; Cellino et al.\ \cite{cellino01}).
 Additionally, since its orbit makes it especially
 accessible to spacecraft and it has been identified as a
 potentially hazardous asteroid (Milani et al.\ \cite{milani09}),
 it was considered an ideal choice for NASA's OSIRIS-REx sample
 return mission (Lauretta et al.\ \cite{lauretta12}). 

 Radar images, taken during its previous two closest approaches in
 1999 and in 2005 (and the next occurred in September 2011 when
 our observations were performed) found 1999~RQ$_{36}$ to be an
 irregular spheroid about 580\,m in diameter (Nolan et al.\
 \cite{nolan07}). The nearly spheroidal shape suggests that it
 might be a strengthless rubble pile, formed by mechanisms
 similar to near-Earth binary systems (Walsh et al.\ \cite{walsh08}).
 The latest analysis, where all available radar data was combined
 with visible lightcurves (Nolan et al.\ \cite{nolan12}), derived
 a mean diameter of 493 $\pm$ 20\,m (mean equatorial diameter
 of 545 $\pm$ 15\,m) and a ``spinning top" shape similar to
 \object{1999~KW$_{4}$} (Ostro et al.\ \cite{ostro06}),
 but with a less well-defined equatorial ridge.
 
 The object's rotation period is 4.297812 $\pm$ 0.000001\,hours
 (Nolan et al.\ \cite{nolan12}), which was derived from combined radar
 and lightcurve measurements. This period is very similar to the 4.288\,h
 found by Nolan et al.\ (\cite{nolan07}) from radar measurements alone 
 and to the 4.2968 $\pm$ 0.0017\,h based on the analysis of lightcurves
 (Hergenrother et al.\ \cite{hergenrother12}) and about
 twice the value given in Krugly et al.\ (\cite{krugly02}) derived
 from incomplete R-band lightcurve observations during the 1999 opposition.
 The rotation period suggests that it has not been greatly
 spun up by tidal or radiation forces. However, the radar-derived pole orientation
 ($\beta_{pole}$ = -90$^{\circ}$ $\pm$ 15$^{\circ}$)
 is the equilibrium state for the YORP effect
 (Vokrouhlick\'y et al.\ \cite{vokrouhlicky03}). 
 The negative pole orientation indicates a retrograde rotation.

 Emery et al.\ (\cite{emery10}) find a size of 610\,m and
 estimate its thermal inertia to be around 600\,Jm$^{-2}$s$^{-0.5}$K$^{-1}$,
 assuming a moderate surface roughness ({\it r.m.s.}-slopes $\sim$20$^{\circ}$).
 Their results are based on Spitzer-IRS spectra and a dual-band
 thermal lightcurve, taken on May 3/4, 2007
 at a phase angle of about 63$^{\circ}$.

 In this paper, we first present our PACS and ESO-VISIR Director
 Discretionary awarded Time (DDT) observations taken of 1999~RQ$_{36}$
 and their data reduction. We follow this with the description of the
 Spitzer-IRS measurements and the corresponding data reduction.
 We analyse the derived flux densities by means of our thermal model.
 We proceed to present our radiometrically derived
 properties (Section \ref{sec:tpm}). In Section~\ref{sec:dis} we
 discuss the object's size, shape, and albedo in a wider context and
 study the influences of surface roughness and thermal inertia in
 detail. We finally conclude the paper with a summary of all 
 derived properties and with the implications of our results for
 the preparations for the OSIRIS-REx mission.

\section{Observations of 1999~RQ$_{36}$}
\label{sec:obs}

\subsection{Herschel-PACS observations and data processing}

 The European Space Agency's (ESA) Herschel Space Observatory
 (Pilbratt et al.\ \cite{pilbratt10}), launched in 2009, performs
 observations from the 2$^{nd}$ Lagrangian point (L2) at
 1.5$\cdot$10$^{6}$\,km from Earth. It has three science
 instruments on board covering the far-infrared part of the
 spectrum, of which the PACS Photometer (covering the wavelength
 range of 60-210\,$\mu$m) has been used to observe 1999~RQ$_{36}$.

 The PACS photometer (Poglitsch et al.\ \cite{poglitsch10})
 imaged the asteroid on September 9, 2011
 (between 19:00 and 21:00 UT) at a heliocentric distance
 of 1.0146-1.0144\,AU and a Herschel-centric distance of
 0.1742\,AU, and the phase angle was 85.3-85.4$^{\circ}$.
 The time used for the observation was DDT-awarded specifically due to
 this optimum closest approach to Earth. The 70/160\,$\mu$m 
 scan-map observation was repeated ten times (5 times at scanangle 70$^{\circ}$ and 5 times
 at 110$^{\circ}$), the 100/160\,$\mu$m observation was
 repeated 14 times (7 times in each scanangle).

 The PACS scan-map measurements were processed using HIPE\,7.0 (Ott et al.\
 \cite{ott09}). The reduction steps were very similar to those
 used for trans-Neptunian objects (Santos-Sanz et al.\ \cite{santos12};
 Mommert et al.\ \cite{mommert12}, Vilenius et al.\ \cite{vilenius12}),
 including corrections for moving targets.
 The final, single-repetition PACS maps (10 at 70\,$\mu$m,
 14 at 100\,$\mu$m, and 24 at 160\,$\mu$m) have been median-\-averaged to
 remove (or at least decrease) the otherwise large instrumental
 noise. The background confusion noise has been eliminated as
 much as possible via techniques that have been developed for other moving
 sources (see detailed description in the electronic appendix in
 Santos-Sanz et al.\ \cite{santos12}). The derived fluxes were aperture-
 and colour-\-corrected to obtain monochromatic flux densities at the PACS
 reference wavelengths. The colour correction values for 1999~RQ$_{36}$ of
 1.01, 1.02, 1.07 in blue (70\,$\mu$m), green (100\,$\mu$m) and red
 (160\,$\mu$m) bands are based on a thermophysical model SED
 (spectral energy distribution), corresponding roughly to a 250\,K
 black-body curve (PACS photometer passbands and colour correction
 factors, \cite{pacs11b}). The flux calibration was verified by a set of
 five high-quality fiducial stars ($\beta$\,And, $\alpha$\,Cet,
 $\alpha$\,Tau, $\alpha$\,Boo, and $\gamma$\,Dra), which were
 observed multiple times in the same PACS observing mode
 (PACS photometer - point source flux calibration, \cite{pacs11a}).

\begin{figure}[h!tb]
 \rotatebox{0}{\resizebox{!}{2.8cm}{\includegraphics{f1a.eps}}}
 \rotatebox{0}{\resizebox{!}{2.8cm}{\includegraphics{f1b.eps}}}
 \rotatebox{0}{\resizebox{!}{2.8cm}{\includegraphics{f1c.eps}}}


 \rotatebox{0}{\resizebox{\hsize}{!}{\includegraphics{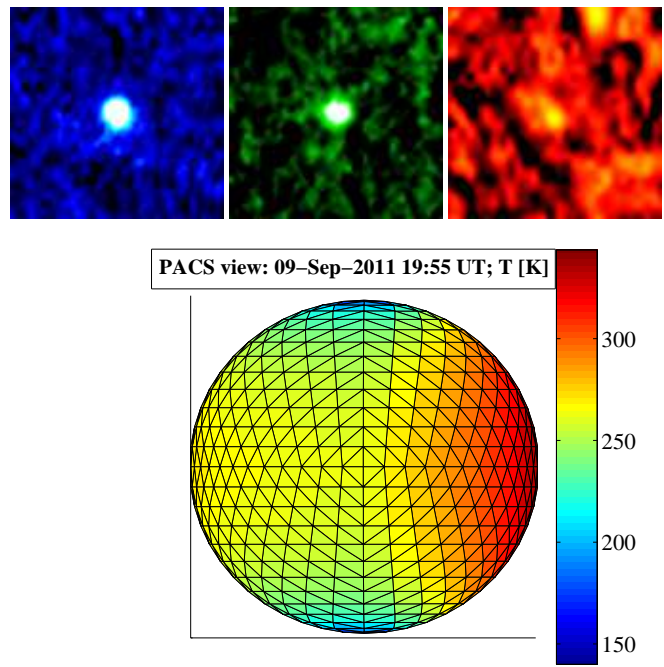}}}
  \caption{The combined PACS images (top), at 70
           (blue), 100 (green), and 160\,$\mu$m (red), along
           with the output from our thermal model (bottom) of
           1999~RQ$_{36}$, at the geometry of Herschel's viewpoint.
           The shorter wavelengths effectively sampled the warmer
           facets of 1999~RQ$_{36}$'s surface, located in the subsolar
           region.
           Note that Herschel cannot spatially resolve the target.
     \label{fig:fig_1}}
\end{figure}

 The final values obtained for the PACS instrument datasets were
 deemed to be accurate within 5\% based upon existing
 calibrations. The absolute flux calibration uncertainty is 
 included in the values given in Table~\ref{tbl:tbl1}
 (added quadratically to observational errors and
 errors related to aperture photometry).
 The fluxes given in Table~\ref{tbl:tbl1} are colour-corrected,
 monochromatic flux densities at 70.0, 100.0, and 160.0\,$\mu$m,
 the three PACS reference wavelengths for the blue, green, and
 red bands, respectively.
 Figure~\ref{fig:fig_1} shows the combined PACS
 images aligned with the output from our thermal model of what
 PACS has observed during its observation campaign.

\begin{table*}
     \caption{Observation summary.}
     \label{tbl:tbl1}
     \begin{tabular}{lrlrrrrrrl}
\hline
\hline
\noalign{\smallskip}
Start Time UT        & Duration & Julian Date & $\lambda_{ref}$ & FD    & FD$_{err}$   & r$_{helio}$ & $\Delta_{obs}$ & $\alpha$ & Observatory/ \\
yyyy-mm-dd hh:mm:ss  & [sec]    & mid-time    & [$\mu$m]        & [mJy] & [mJy]        & r [AU]      & [AU]           & [deg]    & Instrument \\
\noalign{\smallskip}
\hline
\noalign{\smallskip}
2007-05-03 00:00:14 & Start  & 2454223.50016 &        &      &      & 1.1237961 & 0.5056454 & -63.5 & Spitzer-PUI \\    
...                 & 11$\times$ 254 & 2454224.23335 & 16/22 & \multicolumn{2}{c}{Lightcurve}  & 1.1262212 & 0.5089081 & -63.3 & Spitzer-PUI \\    
2007-05-04 11:15:50 & End    & 2454224.96655 &        &      &      & 1.1286398 & 0.5122203 & -63.1 & Spitzer-PUI \\    
\noalign{\smallskip}
2007-05-03 21:39:13 & 4261 & 2454224.42689 & 5.2-38 & \multicolumn{2}{c}{SED} & 1.1268605 & 0.5097779 & -63.2 & Spitzer-IRS \\    
2007-05-03 23:45:57 & 4260 & 2454224.51489 & 5.2-38 & \multicolumn{2}{c}{SED} & 1.1271494 & 0.5101723 & -63.2 & Spitzer-IRS \\    
\noalign{\smallskip}
2011-09-09 19:01:12 & 1414 & 2455814.30068 &  70.0  & 27.0 &  1.7 & 1.0146021 & 0.1742221 & +85.3 &  Herschel/PACS \\  
2011-09-09 19:59:50 & 1414 & 2455814.34140 &  70.0  & 24.3 &  1.5 & 1.0144716 & 0.1742140 & +85.4 &  Herschel/PACS \\  
2011-09-09 19:25:49 & 1978 & 2455814.32104 & 100.0  & 14.2 &  1.1 & 1.0145368 & 0.1742180 & +85.4 &  Herschel/PACS \\  
2011-09-09 20:24:27 & 1978 & 2455814.36176 & 100.0  & 12.0 &  1.2 & 1.0144064 & 0.1742101 & +85.4 &  Herschel/PACS \\  
2011-09-09 19:25:49 & 1978 & 2455814.32104 & 160.0  &  6.5 &  2.3 & 1.0145368 & 0.1742180 & +85.4 &  Herschel/PACS \\  
2011-09-09 19:59:50 & 1414 & 2455814.34140 & 160.0  &  4.4 &  1.9 & 1.0144716 & 0.1742140 & +85.4 &  Herschel/PACS \\  
2011-09-09 20:24:27 & 1978 & 2455814.36176 & 160.0  &  7.6 &  2.0 & 1.0144064 & 0.1742101 & +85.4 &  Herschel/PACS \\  
\noalign{\smallskip}
2011-09-17 09:21:59 & 4$\times$ $\sim$2\,min & 2455821.89306 &   8.59 & 23.7 &  3.6 & 0.9908803 & 0.1798955 & +89.4 &  ESO VLT/VISIR \\   
\noalign{\smallskip}
\hline
     \end{tabular}
\tablefoot{The eleven Spitzer-IRS Peak-up Images (PUI) have the AORKEYs
      2141\,2864/\-4400/\-4912/\-3120/\-5168/\-8240/\-3632/\-4144/\-4656/\-3888/\-3376.
      The data contain simulataneous 16 and 22\,$\mu$m measurements.
      The re-reduced Spitzer-IRS data have the AORKEYs 21412352 and 21412608.
      They are separated by about half a rotation period. See also
      Emery et al.\ (\cite{emery10}) for more observational details and the
      presentation of preliminary results.
      The Herschel-PACS observations have the observation identifiers
      1342228379 ... 1342228382, taken on Operational Day 849, the
      ESO VLT-VISIR observations are related to the programme-ID 287.C-5045(A).}
\end{table*}

\subsection{ESO-VISIR observations and data processing}

 Besides the DDT awarded for Herschel, we also were awarded DDT
 to observe 1999~RQ$_{36}$ in September 2011 via ground-based
 N- and Q-band observations with the ESO-VISIR instrument
 (Lagage et al.\ \cite{lagage04}) mounted on the 8.2\,m VLT
 telescope MELIPAL (UT~3) on Paranal.

 The service-mode observers attempted twice to execute our four 
 observing blocks in imaging mode, including each time the PAH1
 (8.59\,$\mu$m), NeII (12.81\,$\mu$m), and the Q2 (18.72\,$\mu$m)
 filters. The 1999~RQ$_{36}$ observations were done in 
 parallel nod-chop mode with a throw of 8$^{\prime \prime}$.
 Owing to poor observing conditions and
 difficulties during the target acquisition phase, combined with
 end-of-night constraints, the science observations were never
 executed. Nevertheless, we found 1999\,RQ$_{36}$, with
 both the positive and two negative nod-chop beams on-chip,
 in the PAH1 target acquisition images taken on September 17,
 2011 at an airmass of 1.79.
 We processed the four useful raw images using the ESO VISIR
 pipeline {\it Gasgano}\footnote{\tt http://www.eso.org/gasgano}
 to produce a combined image suitable for aperture photometry.
 It was possible to calibrate those data
 using the standard star HD\,217902 (9.923\,Jy at 8.59\,$\mu$m,
 taken at airmass 1.30) that was observed before the acquisition
 of 1999\,RQ$_{36}$ and another one, HD\,16815 (11.679\,Jy at
 8.59\,$\mu$m, taken at airmass 1.40), which was observed the day
 after in the same filter and observing mode. Based on the
 derived count-to-Jy conversion factors and the experience from
 large calibration programmes (Sch\"utz \& Sterzik \cite{schuetz05}),
 we were also able to correct for the airmass difference between
 the calibration stars and our science target. The counts
 decrease by about 10\% for an airmass increase of 0.5.
 The fluxes of the standard stars and our target were extracted
 in an identical way to standard aperture photometry techniques.
 In the case of 1999\,RQ$_{36}$ we checked that the first nod-chop
 pair was consistent with the second (i.e.\ that all three beams
 were in all the acquisition images) by also reducing them separately.
 The derived fluxes from each pair were consistent with those
 produced by combining both pairs together. We applied a
 colour correction of 1.004 (for 150-250\,K blackbodies, considering
 the full atmosphere
 and bandpass transmission profiles) to the measured 1999\,RQ$_{36}$
 flux to obtain a mono-chromatic flux density at the PAH1
 reference wavelength (see Table~\ref{tbl:tbl1}). For the error 
 calculation (E.\ Pantin, M.\ van den Ancker \& R.\ Siebenmorgen, priv.\ comm.),
 we quadratically added the following error sources:
 error in count-to-Jy conversion factor (2\%), error of stellar
 model (2\%), photometry error (10\%), error of colour correction
 (2\%), estimated additional uncertainty for measurements at
 airmass larger than 1.5 (5\%), and flat-field error for an off-centre
 position in the acquisition images (10\%), summing up to a total 
 of 15\% error in the derived flux density.

\subsection{Observations of 1999 RQ36 with the Spitzer Space Telescope}

 To complete the dataset, we also analysed and utilised data
 from the Spitzer Space Telescope (Werner et al.\ \cite{werner04}).
 Emery et al.\ (\cite{emery10}) present part of these observations,
 but the results were based on a preliminary data reduction alone.

 Spitzer observed 1999~RQ$_{36}$ in different modes.
 Thermal IR spectra from 5.2 to 38\,$\mu$m were taken
 with the Infrared Spectrograph (IRS) of opposite hemispheres of the body
 on 3/4 May 2007 21:42:03 - 22:46:57 (ID 21412352) and
 23:48:51 - 00:53:40 UT (ID 21412608).
 Photometry at 16 and 22\,$\mu$m was obtained with the IRS peak-up imaging
 (PUI) mode (IDs 21412864/\-4400/\-4912/\-3120/\-5168/\-8240/\-3632/\-4144/\-4656/\-3888/\-3376),
 taken at regular intervals between 3 May 2007 00:00:15 UT and
 4 May 2007 11:11:37 UT. The 11 measurements cover the full rotation
 period uniformly. The phase angles, the heliocentric, and
 Spitzer-centric distances are given in Table~\ref{tbl:tbl1}.

\begin{figure}[h!tb]
 \rotatebox{90}{\resizebox{!}{\hsize}{\includegraphics*{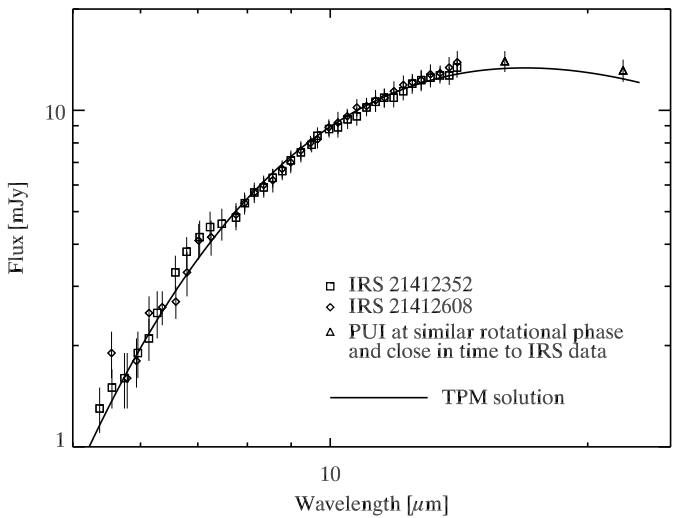}}}
  \caption{Both Spitzer IRS spectra (SL-part only) are shown
           together with 2 PUI measurements taken close in
           time at similar rotational phases. Our best TPM
           solution is shown as a solid line.
     \label{fig:fig_2}}
\end{figure}

 The {\em Spitzer}-IRS observations were reduced with
 SPICE\footnote{{\tt http://irsa.ipac.caltech.edu/\-data/\-SPITZER/\-docs/\-dataanalysistools/\-tools/\-spice/} [May 7$^{th}$, 2012]}
 in the recommended way following the data analysis cookbook for moving
 targets\footnote{{\tt http://irsa.ipac.caltech.edu/\-data/\-SPITZER/\-docs/\-dataanalysistools/\-cookbook/1} [May 7$^{th}$, 2012]}.
 We used the PBCD-products\footnote{Post-Basic Calibrated Data: Level 2 data} from the Spitzer Heritage
 Archive SHA\footnote{{\tt http://sha.ipac.caltech.edu/\-applications/\-Spitzer/\-SHA/} [May 7$^{th}$, 2012]},
 subtracted the background measurements from the on-source
 measurements, and produced two spectra for each order in
 the short-low (SL) and long-low (LL) modules\footnote{Note, that we used only 
 the SL-part of the observations. The LL part is more noisy and does not constrain
 our analysis in a significant way.}, using the appropriate
 uncertainty and mask files. We averaged the resulting A-B and B-A spectra.
 The low flux in the SL2 range (5.2 to 7.5\,$\mu$m) was averaged using a boxcar filter;
 in addition, we resampled the data in bins of uniform resolution
 R = $\lambda/ \Delta \lambda$, with R = 40 (SL1: 7.6 to 14.3\,$\mu$m)
 and R = 30 (SL2) to better measure the
 continuum signal of 1999~RQ$_{36}$ and to reduce the number of data points
 for the thermophysical model (TPM) analysis. For each bin we determined the weighted
 averaged wavelength, the median flux, and the root-sum-square of the propagated
 errors, including a 5\% absolute flux calibration error
 as specified by the {\em Spitzer}/IRS team. Our derived
 fluxes and errors agreed within the given errorbars with
 the published figures in Emery et al.\ (\cite{emery10}) and
 are shown in Fig.~\ref{fig:fig_2}.

\subsection{Observation summary}

 In total, including the Herschel, ESO-VISIR and Spitzer dataset
 (see Table~\ref{tbl:tbl1}), we were able to place into our model a
 set of observations covering the wavelengths from 6 to 160\,$\mu$m,
 phase angles of -63$^{\circ}$, +85$^{\circ}$, and +89$^{\circ}$
 (positive before opposition, negative
 after opposition), and covering nicely all rotational phases.
 For the TPM analysis (Sect.~\ref{sec:tpm}) we use the true
 observing and illumination geometries (Spitzer-, Herschel-, VLT-centric)
 for each measurement individually as given in Table~\ref{tbl:tbl1}.
 The importance of this dataset becomes clear in the coming section
 where it can be seen that our comprehensive coverage of wavelength and
 phase angle is key to extracting the main thermal parameters output
 from our thermal model.

\section{Thermophysical analysis of the observational data}
\label{sec:tpm}

 We started the thermophysical model (TPM) analysis of the complete dataset
 by using a simple spherical shape model. For the object's
 rotation period we used 4.297812 $\pm$ 0.000001\,hours (Nolan et al.\ \cite{nolan12}).
 Thermal data are very sensitive to the orientation of the spin axis
 (see e.g., M\"uller et al.\ \cite{mueller11}) and we used it as a
 free parameter in our analysis. Due to computing speed limitations we restricted
 this exercise to 60 different orientations for the rotation axis
 distributed over the entire sphere. The ecliptic latitude
 and longitude pairs to describe the spin vector orientation
 correspond to the 60 vertices of a mathematical description of a
 truncated icosahedron\footnote{\tt http://en.wikipedia.org/wiki/Truncated\_icosahedron}
 with 12 regular pentagonal faces and 20 regular hexagonal faces.
 To complete the picture, we added a few key spin axis orientations:
 pro- and retrograde rotations with the spin vector perpendicular
 to the ecliptic plane ($\beta_{pole}$ = $\pm$90$^{\circ}$), pole-on
 viewing geometries as seen during the Spitzer, the Herschel, and the
 VISIR observations, and another four orientations (with $\beta_{pole} < -70^{\circ}$)
 to explore the YORP spin-up equilibrium state (Vokrouhlick\'y et
 al.\ \cite{vokrouhlicky03}), which corresponds to the latest
 value derived by Nolan et al.\ (\cite{nolan12}): The obliquity
 180$^{\circ}$ case for the spin vector ($\beta_{pole}$ = -83.98$^{\circ}$,
 $\lambda_{pole}$ = 92.17$^{\circ}$) is very well covered by by this set
 of different spin axis orientations, which included values of -78$^{\circ}$,
 -80$^{\circ}$, and -90$^{\circ}$ for $\beta_{pole}$ (each time for a range
 of $\lambda_{pole}$ values).
 
 The spherical shape model with the almost 70 different spin axis
 orientations were imported into our TPM code
 (Lagerros \cite{lagerros96}; \cite{lagerros97}; \cite{lagerros98}).
 Details about the techniques have recently been described in
 M\"uller et al.\ (\cite{mueller11}) and O'Rourke et al.\ (\cite{orourke12}).
 The TPM produces accurate thermal IR spectra and thermal lightcurves,
 taking a number of physical and thermal processes into account.
 In the TPM, the object is described by a given size, shape, spin-state,
 and albedo, placed at the true observing and illumination geometry.
 The TPM considers a 1-d heat conduction into the surface and
 allows surface roughness to be included, described by
 ``f", the fraction of the surface covered by spherical crater
 segments and ``$\rho$", the r.m.s.\ of the surface slopes, connected
 to the crater width-to-depth ratio (Lagerros \cite{lagerros98}).
 A default setting (M\"uller et al.\ \cite{mueller99}) with
 $\rho$=0.7 and f=0.6 has been used as baseline.

 The contributions of the subsurface emission at longer wavelength is
 accounted for by a wavelength-dependent emissivity decreasing from 0.9
 at mid/far-IR (5 to 100\,$\mu$m) to about 0.8 in the
 submillimetre/millimetre range of the spectrum, derived from a
 combined set of large main-belt asteroids (M\"uller and
 Lagerros \cite{mueller98}; \cite{mueller02}). Here, all radiometrically
 critical measurements are taken at wavelength below 100\,$\mu$m, and
 there is no difference in using a constant emissivity of 0.9 or
 the wavelength-dependent emissivity model presented in M\"uller \&
 Lagerros (\cite{mueller02}).

 For the full treatment of the energy balance on each surface facet
 within the model, we need to describe the amount of reflected light,
 described by the commonly used H-G-system (Bowell et al.\ \cite{bowell89}).
 The value for the absolute V-band magnitude H$_V$ was determined
 by Krugly et al.\ (\cite{krugly02}) and recently refined by
 Hergenrother et al.\ (\cite{hergenrother12}). The best value of
 H$_V$ = 20.65\,mag was obtained by a linear phase function fit
 (phase slope $\beta$= 0.039$\pm$0.005\,mag) to V-band photometry
 covering the phase angle range from 15$^{\circ}$ to 100$^{\circ}$.
 Hergenrother et al.\ (\cite{hergenrother12}) discuss the issue
 of a shallow opposition effect for low-albedo asteroids and conclude
 that a more realistic H$_V$-value for 1999~RQ$_{36}$ would be
 20.5$\pm$0.3\,mag. For G (slope parameter) we used the Krugly
 et al.\ (\cite{krugly02}) value of 0.12.

\begin{figure}[h!tb]
 \rotatebox{90}{\resizebox{!}{\hsize}{\includegraphics*{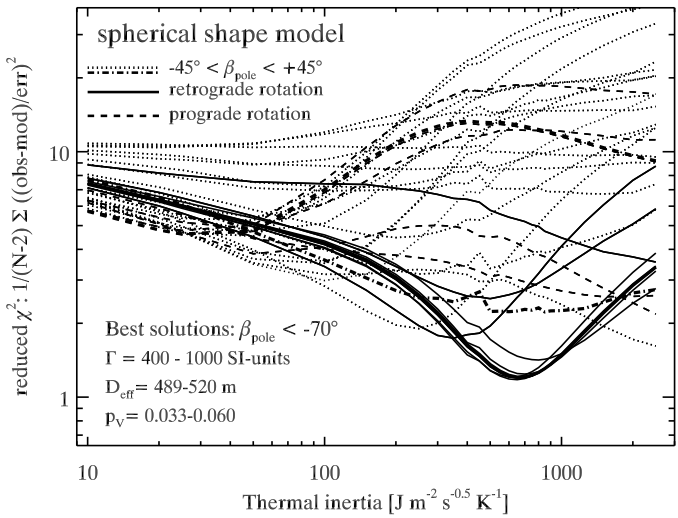}}}
  \caption{$\chi^2$ test for the full set of spin vector orientations and
           the full range of thermal inertias. The retrograde solutions
           with $\beta_{pole} < -45^{\circ}$ are given as solid lines,
           the prograde ones with $\beta_{pole} > +45^{\circ}$
           as dashed lines, the ones where the latitude of the rotation
           pole is in the range $-45^{\circ} < \beta_{pole} < +45^{\circ}$
           are shown in dotted lines. The thicker solid and dashed
           lines represent the $\beta_{pole} = \pm90^{\circ}$ solutions,
           the thick dashed-dotted line the Herschel pole-on geometry.
     \label{fig:fig_3}}
\end{figure}

 Based on all available thermal data we searched for the best size and
 albedo solutions via well-established radiometric techniques
 (e.g., Harris \& Lagerros \cite{harris02} and references therein)
 by using the above described settings. For each of the above
 spin vector orientations we calculated the reduced $\chi^2$-value
 for a wide range of typical thermal inertias (Delbo et al.\
 \cite{delbo07}). We calculated the the reduced $\chi^2$-values
 in the following way:
 $\chi^2_{reduced}$ = $1/(N-2) * \sum ((flx_{obs} - flx_{mod})/err_{obs})^2$,
 with N the number of degrees of freedom.
 Figure~\ref{fig:fig_3} shows the result of the
 analysis. The lowest $\chi^2$ values are found for retrograde
 rotations ($\beta_{pole} < -70^{\circ}$) and thermal
 inertias around 600\,Jm$^{-2}$s$^{-0.5}$K$^{-1}$. The thermal inertia
 3-$\sigma$ confidence interval is from 400 to just above
 1000\,Jm$^{-2}$s$^{-0.5}$K$^{-1}$. The 3-$\sigma$ confidence interval
 $\chi^2 < (\chi^2_{min} + 3^2$) is based on actual $\chi^2$ = $\sum{((obs-mod)/err)^2}$
 calculations (not the reduced $\chi^2$ as shown in Fig.~\ref{fig:fig_3}).

 Considering this possible $\Gamma$-range in combination with a spherical
 shape model and a pole axis pointing to the ecliptic south pole ($\beta_{pole} = -90^{\circ}$,
 retrograde rotation) leads to a size in the range 498 to 513\,m
 and a geometric albedo p$_V$ in the range 0.043 to 0.046. The true errors are
 larger: (i) the TPM fit to the observations is not perfect (mainly due to the
 spherical shape model), and the {\it r.m.s.} of
 all observation/model ratios (see Figs.~\ref{fig:fig_4} \& \ref{fig:fig_5})
 has to be included (dominating error contribution for the diameter);
 (ii) the solution also depends slightly on the selected spin vector
 orientation within the possible range $\beta_{pole} < -70^{\circ}$;
 (iii) the H-magnitude is given with an error of $\pm$0.3\,mag
 (dominating error contribution for the albedo).
 Combining all errors quadratically led to an effective size of 503$^{+17}_{-14}$\,m
 for a spherical shape and a geometric albedo of 0.045$^{+0.015}_{-0.012}$.
 These numbers are based on the assumption that 1999~RQ$_{36}$ has
 a ``default roughness" that is typical of large main-belt asteroids
 (see also discussion in Section~\ref{sec:dis} below).

\section{Discussions}
\label{sec:dis}

\subsection{Spin-axis orientation, sense of rotation}
\label{sec:spin}

 The before/after opposition data are the key element for extracting the sense
 of rotation. Our Herschel observations are all taken before opposition,
 i.e.\ leading the Sun (here positive phase angles). This means that 
 Herschel has observed the target, while the unilluminated part of
 the surface was cold (see Fig.~\ref{fig:fig_1}) corresponding to the
 pre-dawn part of the surface (for a retrograde sense of rotation).
 Spitzer has seen 1999~RQ$_{36}$ after opposition when the unilluminated
 part was still warm after it had rotated out of the Sun (see Fig.~\ref{fig:fig_7}).
 The importance
 of the sense of rotation can clearly be seen in Fig.~\ref{fig:fig_3}. The
 prograde rotation options produced very high values for the reduced $\chi^2$
 and can therefore be excluded with high confidence. The models with
 spin axes close to the ecliptic plane (including pole-on geometries)
 constrain the thermal inertia only very little; nevertheless, the reduced
 $\chi^2$-values are very high so these solutions can be ruled out.
 The TPM analysis clearly favours an axis orientation with
 $\beta_{pole} < -70^{\circ}$ in the ecliptic coordinate system.
 This agrees with the findings
 by Nolan et al.\ (\cite{nolan12}) that the pole is close to the equilibrium
 state for the YORP effect (Vokrouhlick\'y et al.\ \cite{vokrouhlicky03}).
 1999~RQ$_{36}$ has an orbit inclination of 6.02$^{\circ}$, and $\Omega$, the
 argument of ascending node, is 2.17$^{\circ}$. The YORP predicted asymptotic zone
 with obliquity $\epsilon$ of 180$^{\circ}$ would therefore correspond to
 $\beta_{pole}$ = -83.98$^{\circ}$ and $\lambda_{pole}$ = 92.17$^{\circ}$
 in our convention, very close to the south ecliptic pole.

\begin{figure}[h!tb]
 \rotatebox{90}{\resizebox{!}{\hsize}{\includegraphics*{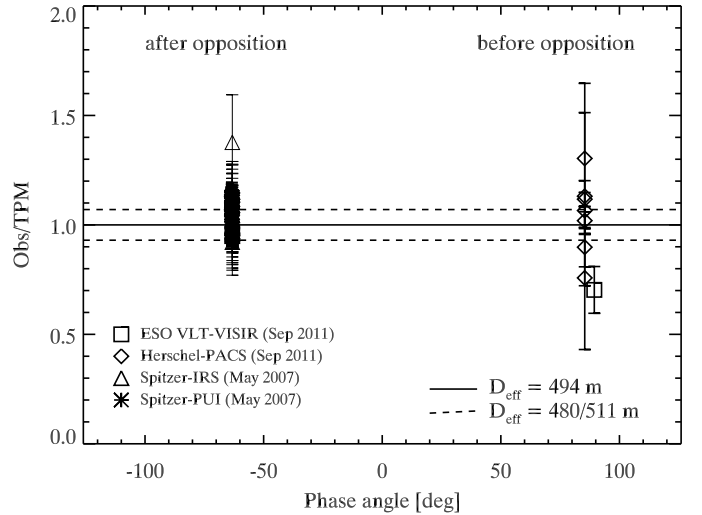}}}
  \caption{Ratio between the observed fluxes and the corresponding
           TPM predictions as a function of phase angle (negative: after opposition;
           positive: before opposition). The model calculations are based on
           a thermal inertia of 600\,Jm$^{-2}$s$^{-0.5}$K$^{-1}$ and $\rho$=0.2 for
           the surface roughness. The VISIR data point can only be matched by assuming
           a higher surface roughness (see also Section~\ref{sec:roughness} for further
           discussions).
     \label{fig:fig_4}}
\end{figure}

\subsection{Surface roughness}
\label{sec:roughness}

\begin{figure}[h!tb]
 \rotatebox{90}{\resizebox{!}{\hsize}{\includegraphics*{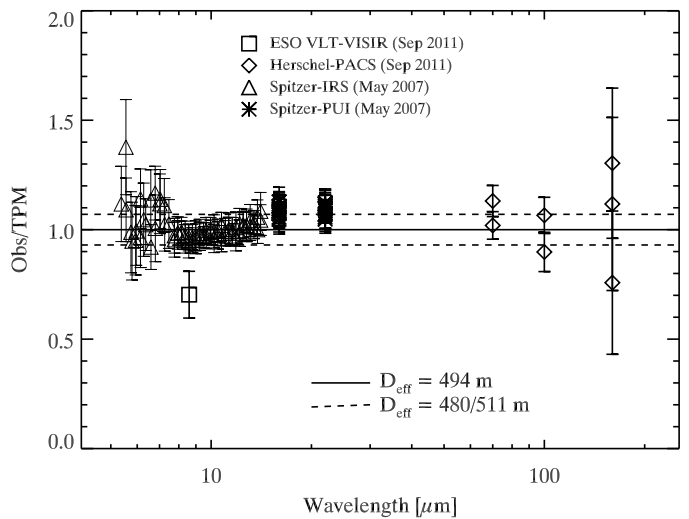}}}
  \caption{Ratio between the observed fluxes and the corresponding
           TPM predictions as a function of wavelength.
     \label{fig:fig_5}}
\end{figure}

 Due to the lack of observations close to opposition, our dataset
 does not constrain the surface roughness that plays
 a more dominant role at phase angles close to opposition very well
 (M\"uller \cite{mueller02a}). In Section~\ref{sec:tpm} we used
 the ``default roughness" that was derived for large main-belt asteroids
 that are covered with a low thermal conductivity regolith
 (M\"uller et al.\ \cite{mueller99}). To study the influence
 of roughness on the final size and albedo solutions, we
 analysed all thermal data, but this time for different levels of
 roughness. For simplicity, we assumed that 100\% of the
 surface is covered by craters (f=1.0), which is a reasonable assumption
 considering that the model crater definition also includes micro-craters
 down to very small scales where the geometric optics approximation is
 still valid (M\"uller \& Lagerros \cite{mueller98}).
 We varied the {\it r.m.s.} of the surface slopes $\rho$
 in steps of 0.1 from a perfectly smooth surface ($\rho$=0.0) to
 a very rough surface ($\rho$=1.0) and repeated our radiometric analysis.
 The reduced $\chi^2$-minima indeed reach acceptable values
 for almost all levels of roughness:
 A very low surface roughness with $\rho$ $<$ 0.1 would lead to a
 thermal inertia of 450\,Jm$^{-2}$s$^{-0.5}$K$^{-1}$ which corresponds to
 an effective size of 490\,m and a geometric albedo of p$_V$ = 0.047.
 A surface roughness with $\rho$ $>$ 0.9 would
 point to a thermal inertia of 850\,Jm$^{-2}$s$^{-0.5}$K$^{-1}$ with
 D$_{eff}$ = 508\,m and p$_V$ = 0.044.
 This size range is entirely consistent with the possible range given 
 by radar (Nolan et al.\ \cite{nolan12}),
 but it confirms that our dataset suffers from the degeneracy between surface
 roughness and thermal inertia.
 Figures~\ref{fig:fig_4} and \ref{fig:fig_5} show the comparison between
 observations and model predictions assuming a roughness of $\rho$=0.2
 (with f=1.0), similar to the moderate surface roughness (r.m.s.\ slopes $\sim$20$^{\circ}$)
 used by Emery et al.\ (\cite{emery10}) to fit the IRS data.

\begin{figure}[h!tb]
 \rotatebox{90}{\resizebox{!}{\hsize}{\includegraphics*{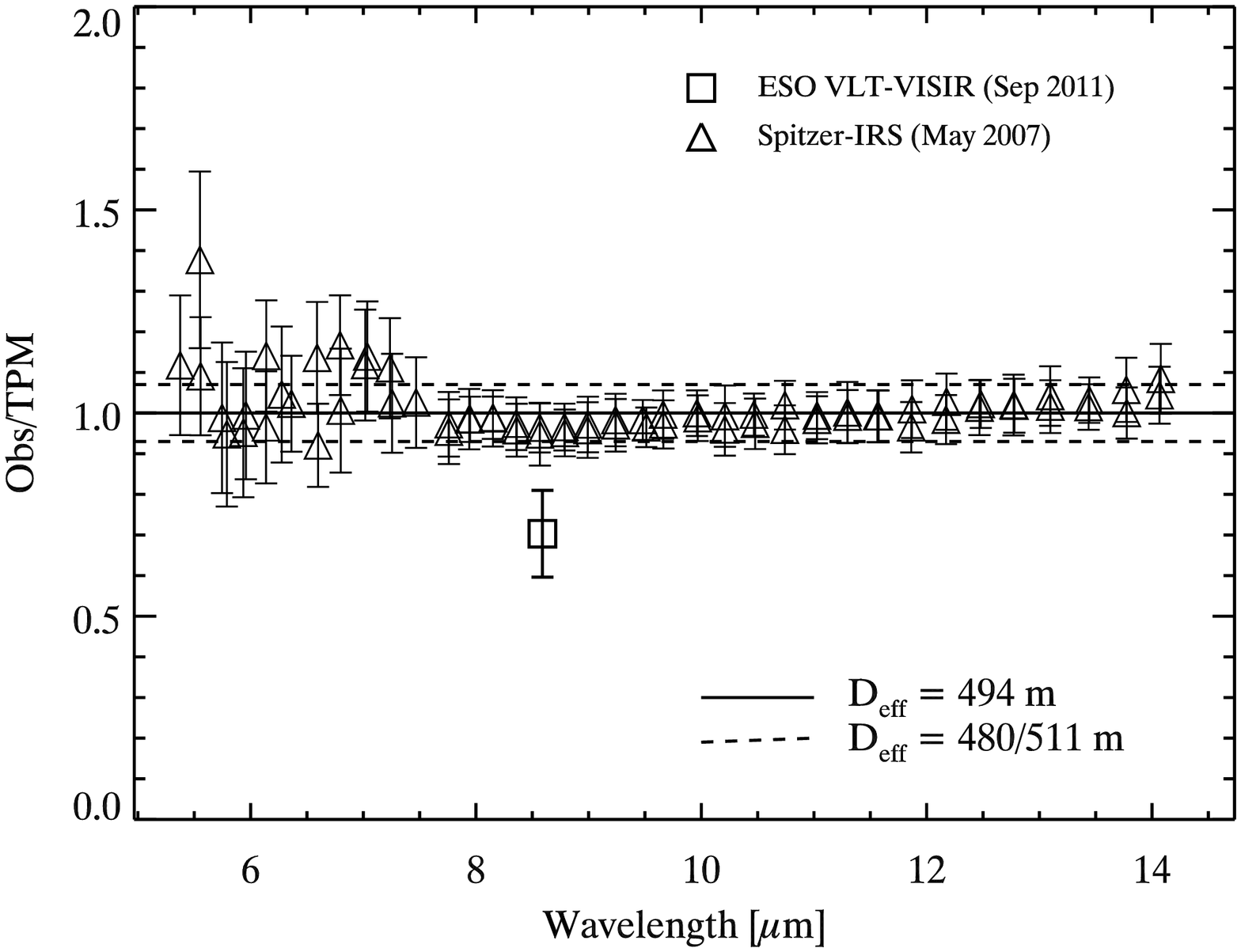}}}
 \rotatebox{90}{\resizebox{!}{\hsize}{\includegraphics*{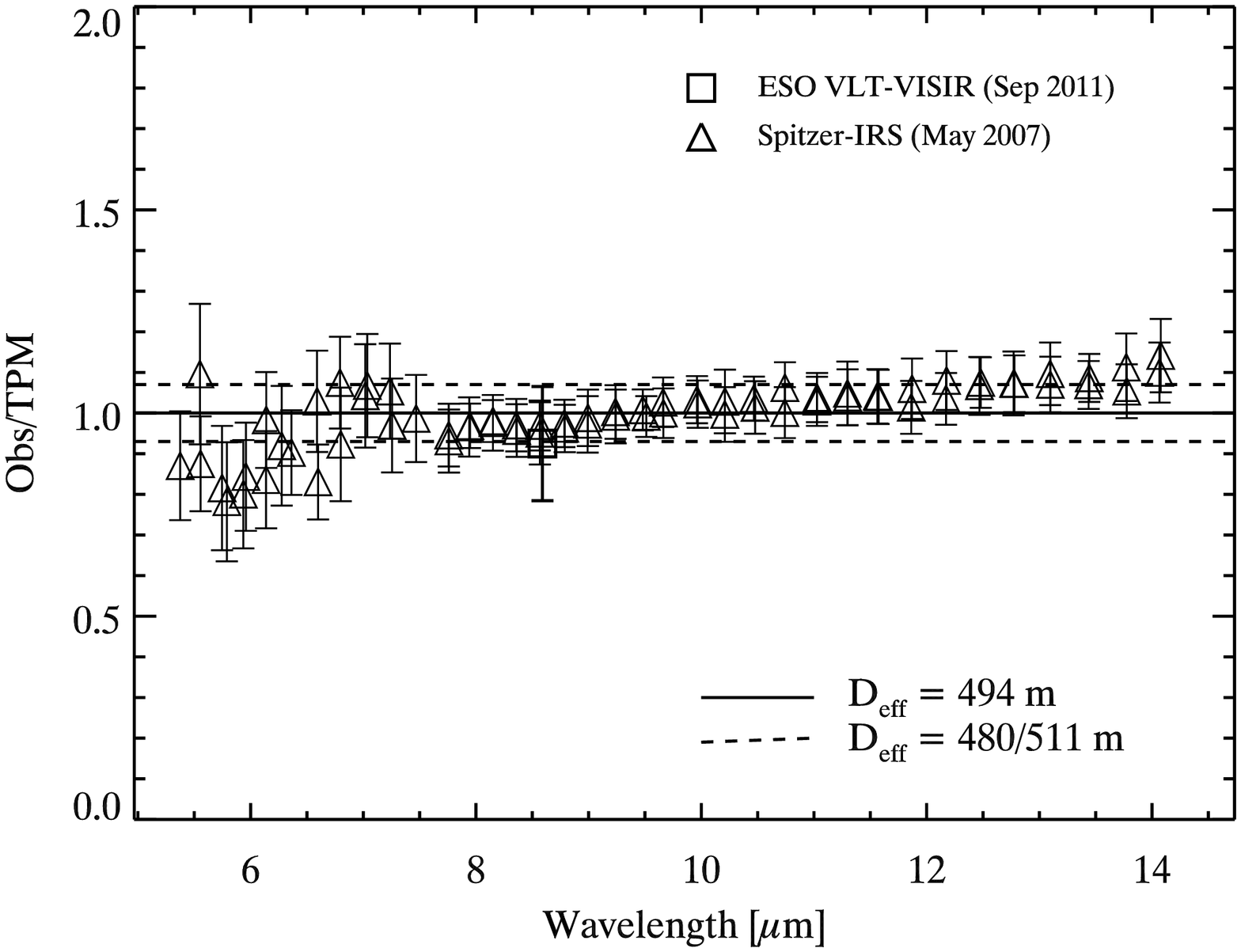}}}
  \caption{Ratio between the observed fluxes and the corresponding
           TPM predictions as a function of wavelength. Top:
           assuming moderate surface roughness with a r.m.s.-slope of 0.2 gives a 
           nice match to the 2 Spitzer-IRS spectra taken at opposite hemispheres
           (reduced $\chi^2$ $<$ 0.75).
           Bottom: assuming a high surface roughness with a r.m.s.-slope $>$ 0.8 gives
           a nice match to the VLT-VISIR data point, but there remains a slight trend with
           wavelengths for the IRS data which are taken at different rotation phases
           (reduced $\chi^2$ $>$ 1.25).
     \label{fig:fig_6}}
\end{figure}

 The only data that do contain some information about roughness are the
 data at short wavelengths below 12\,$\mu$m (IRS-spectrum and VISIR data point).
 The Wien part of the SED is strongly influenced by the hottest temperatures
 in the sub-solar region, while data at longer wavelengths beyond the thermal
 emission peak are more connected to the disk-averaged temperatures.
 This can be seen in Fig.~\ref{fig:fig_6}:
 the low-roughness case fits the IRS-data very well
 (top of Fig.~\ref{fig:fig_6}), but the model prediction for the
 VISIR data-point is overestimated by about 30\%. If we use
 a high-roughness solution, then it is the other way round, where the
 VISIR data-point is well matched by the model prediction, but
 the IRS-spectrum divided by the corresponding model prediction
 shows a slight, wavelength-dependent mismatch (bottom of
 Fig.~\ref{fig:fig_6}). The effects are small, and the VISIR
 data point has been taken in marginal conditions and as an
 acquisition frame. Also, modelling aspects like the implementation
 of surface roughness might play a role.
 The VISIR and IRS dataset are taken at very large and very
 different phase angles where the roughness effects on the IR beaming
 have not been studied very well. Nevertheless, the VISIR data point
 might indicate that we see variations in surface properties while
 the object is rotating. Based on the above specified rotation
 period, the two IRS spectra and the VISIR data are indeed taken
 at different rotational phases and correspond to different
 regions on the surface. A look at the reduced $\chi^2$-values
 seems to confirm this idea. The low-roughness model (top of Fig.~\ref{fig:fig_6})
 matches our thermal dataset very well (reduced $\chi^2$-values below 0.75), while
 the high roughness model (bottom of Fig.~\ref{fig:fig_6})
 clearly has problems with many of our data points (reduced $\chi^2$-values
 above 1.25), although it produces a perfect fit to the VISIR data point, which
 was taken at a unique rotational phase.
 But this aspect of surface roughness variations is clearly on
 vague grounds and should be confirmed via radar measurements or
 short-wavelength thermal lightcurves.

 We should note here that the thermal emission at
 wavelengths shorter than 12$\mu$m originates mainly
 in the hottest sub-solar region, which is equatorial
 in the case of 1999~RQ$_{36}$. Our speculation about
 possible surface roughness heterogeneity is therefore only
 related to a broad equatorial zone. Our dataset is not
 sensitive to roughness variations in the polar regions.

 Roughness can in principle be present on any scale. Radar observations
 from Goldstone and Arecibo in 1999 and 2005 revealed a featureless surface
 down to the radar resolution limit of 7.5\,m (Hudson et al.\ \cite{hudson00};
 Nolan et al.\ \cite{nolan07}), but so far the radar team(s) have not
 investigated whether there are also variations in surface roughness seen in the
 radar echoes (Mike Nolan, private communication).

\subsection{Thermal inertia $\Gamma$}

 Pre- and post-opposition observations are the key to constraining
 the thermal inertia $\Gamma$, even more when the observations are
 taken at large phase angles (see Fig.~\ref{fig:fig_4}).
 But, as described above, uncertainty remains because of
 the degeneracy between surface roughness effects and thermal inertia.
 From the exercise with different levels of roughness we find thermal
 inertias ranging from 450 to 850\,Jm$^{-2}$s$^{-0.5}$K$^{-1}$. If we also
 include variations for the spin pole ($\beta_{pole}$ = -90$^{\circ}$ $\pm$ 20$^{\circ}$),
 the possible confidence range lies between 350 to 950\,Jm$^{-2}$s$^{-0.5}$K$^{-1}$,
 with the most likely value at 650\,Jm$^{-2}$s$^{-0.5}$K$^{-1}$.

 A closer inspection of possible thermal inertias
 confirms that thermal inertias outside this range cause a very obvious
 mismatch between the data taken before opposition (positive phase
 angles in our convention) and the ones taken after opposition(negative
 phase angles): A figure similar to Fig.~\ref{fig:fig_4}, but with
 thermal inertias less than 350 or greater than 950\,Jm$^{-2}$s$^{-0.5}$K$^{-1}$
 would show a large imbalance between data taken before opposition and
 data taken after opposition.
 The determined $\Gamma$-range is close to the 600\,Jm$^{-2}$s$^{-0.5}$K$^{-1}$
 derived by Emery et al.\ (\cite{emery10}) for 1999~RQ$_{36}$ and also
 close to to $\Gamma$ = 750\,Jm$^{-2}$s$^{-0.5}$K$^{-1}$ derived for (25143)~Itokawa
 (M\"uller et al.\ \cite{mueller05}).
 Such a thermal inertia value indicates regions
 with high conductivity, i.e.\ coherent boulders.
 It can therefore be expected
 that the surface of 1999~RQ$_{36}$ might show a similar texture
 to Itokawa's: with zones with different surface rock size
 distributions, though possibly with fewer large boulders
 that would have shown up as surface features
 in the radar measurements (Nolan et al.\ \cite{nolan12}).
 But, as discussed before, our results are
 only connected to the warm region where the measured thermal flux
 originates.
 Other parts of the surface -at very large and very low latitudes-
 could have very different thermal properties. Seismic vibrations or
 YORP-introduced regolith turn-over (see Binzel et al.\  \cite{binzel10},
 and references therein) could have influenced these ``invisible" parts
 by moving material towards the equatorial region that we ``see" in
 our thermal dataset. This means that our analysis does not preclude
 some regions with small grains.

\subsection{Spheroidal or ellipsoidal shape}

 The visual lightcurves presented by Hergenrother et
 al.\ (\cite{hergenrother12}) were taken at large phase angles between
 60$^{\circ}$ and 70$^{\circ}$ and showed a low amplitude of 0.17\,magnitudes
 and a trimodal shape. It is not possible to reproduce the observed
 lightcurve shape with a simple ellipsoidal shape model of uniform albedo,
 but small shape deviations in combination with high phase angles could
 explain the trimodal lightcurve. The lightcurve amplitude
 (at a phase angle of 65$^{\circ}$) can be reproduced by a slightly
 deformed body with an axis ratio of a/b=1.04 and rotating
 around a c-axis that points at the ecliptic south pole. We tested if such a
 rotating ellipsoid in combination with our derived mean thermal inertia
 of 600\,Jm$^{-2}$s$^{-0.5}$K$^{-1}$ could also
 explain the two-band Spitzer-PUI thermal lightcurve (see also Figure 2
 presented in Emery et al.\ \cite{emery10}). Axis ratios a/b above about 1.1
 (for the given spin vector and thermal inertia) can be excluded
 on basis of these 16 and 22\,$\mu$m lightcurves.

\begin{figure}[h!tb]
 \rotatebox{90}{\resizebox{!}{\hsize}{\includegraphics*{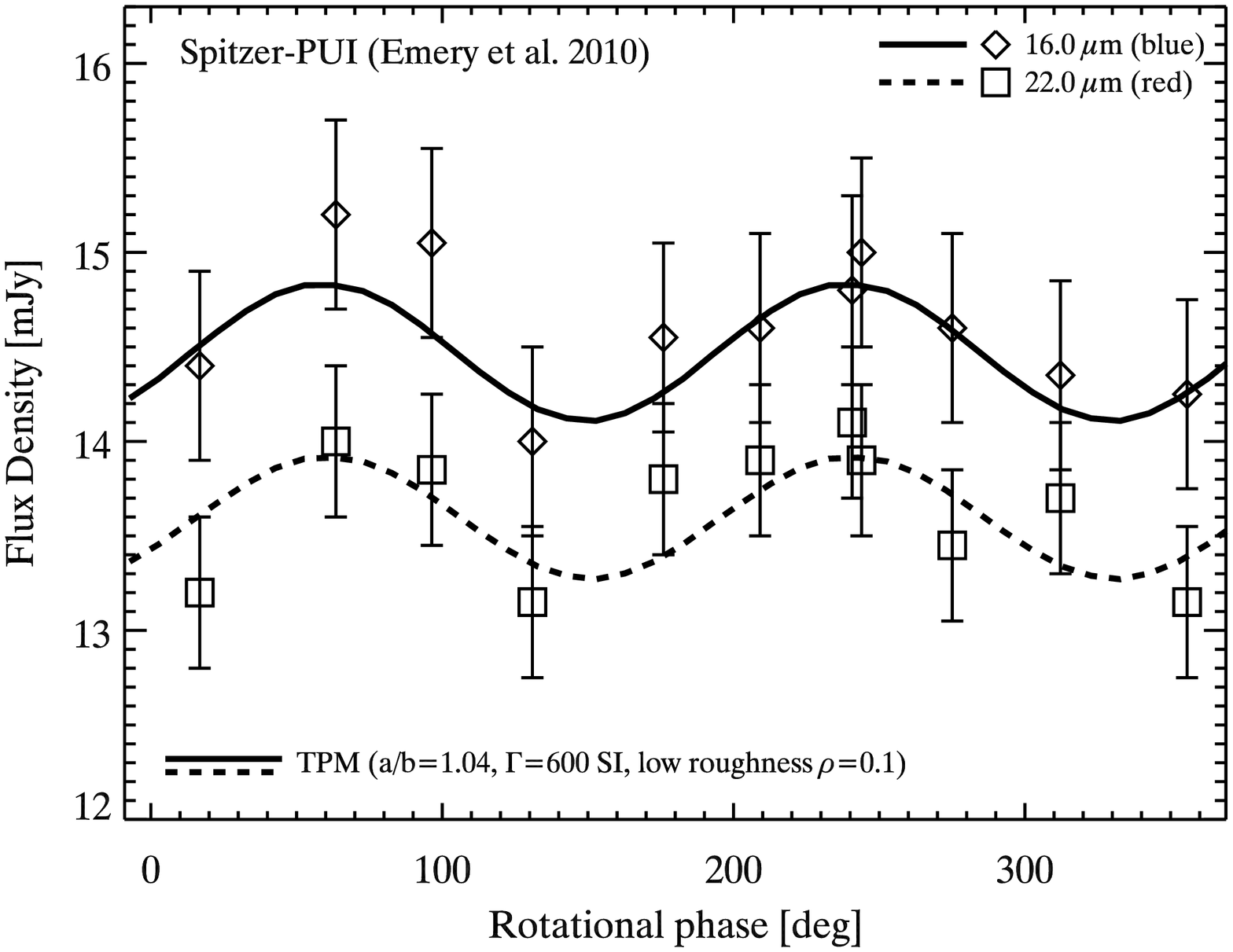}}}
 \rotatebox{0}{\resizebox{\hsize}{!}{\includegraphics{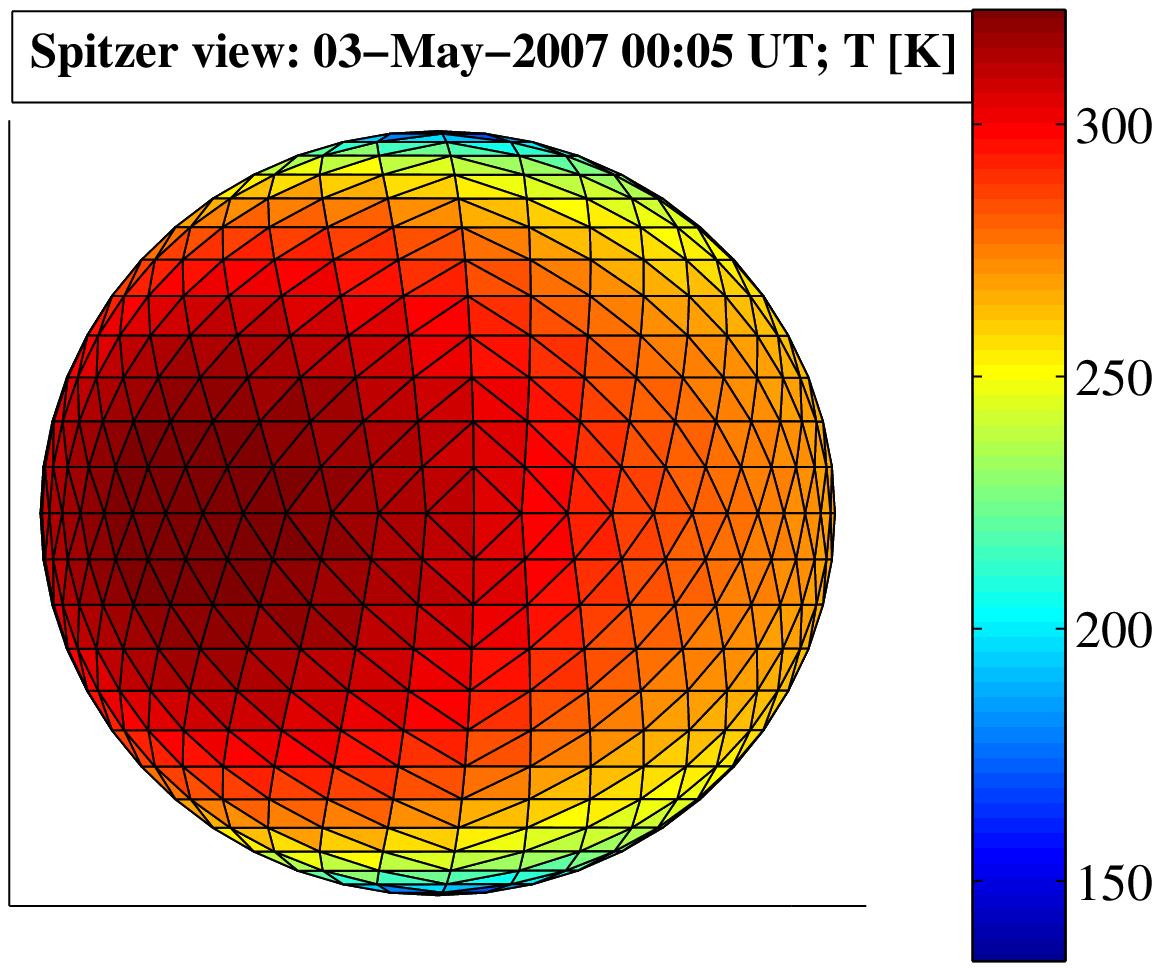}}}
  \caption{Top: Spitzer-PUI data with the TPM prediction overplotted.
           Bottom: The corresponding output from our thermal model of 1999~RQ$_{36}$
           as seen by Spitzer during its observation campaign.
           Spitzer cannot resolve the target spatially.
     \label{fig:fig_7}}
\end{figure}

 Figure~\ref{fig:fig_7} shows the result as a function of rotational
 phase. The zero rotational phase was determined in 10$^{\circ}$
 steps via a $\chi^2$-test against the observational Spitzer-PUI data.
 The thermal lightcurve of a slightly elongated shape model with a
 180$^{\circ}$ obliquity of the spin vector (combined with the derived
 low surface roughness and a thermal inertia of 600\,Jm$^{-2}$s$^{-0.5}$K$^{-1}$)
 explains the different absolute flux levels at 16 and 22\,$\mu$m very well
 and, at the same time, matches the two-band PUI thermal lightcurve and
 amplitude better than a simple spherical shape model.
 A lower thermal inertia would (for the given PUI observing geometry)
 increase the thermal lightcurve amplitude, a larger inertia would lower
 the amplitude. But the uncertainties of the individual PUI measurements
 and the remaining degeneracy with roughness do not allow
 constraining the thermal inertia even more.
 Unfortunately, the {application of our PUI-derived rotating ellipsoidal model
 directly to the PACS dataset was not successful. The PACS data cover about
 1.5\,h of the rotation period, but the} quality of the PACS dataset with
 respect to S/N and time resolution is not sufficient to repeat the thermal
 lightcurve exercise as we did for the PUI data. The predicted 70.0\,$\mu$m
 lightcurve amplitude is about 1\,mJy peak-to-peak, while the flux error of
 the best 70.0\,$\mu$m data point is already 1.5\,mJy.

\subsection{Size}

 The unknowns in surface roughness influenced the optimum thermal
 inertia slightly (see above), but the size and albedo values did not change
 significantly. We repeated the exercise shown in
 Fig.~\ref{fig:fig_3} with a the low surface roughness (r.m.s.-slope $\rho$ = 0.2),
 which produced the best match to the Spitzer-IRS data
 using a spherical and the ellipsoidal shape model specified above.
 The $\chi^2$ values are now a few percent lower due to the better fit
 to the shortest wavelength data and, in the case of the ellipsoidal shape model,
 also a better match to the PUI data.
 The $\chi^2$-minimum is pointing towards a thermal inertia of
 $\Gamma$=650\,Jm$^{-2}$s$^{-0.5}$K$^{-1}$, resulting in an equivalent
 diameter of an equal volume sphere of D$_{eff}$ = 494$^{+11}_{-14}$\,m
 when we consider the full 3-$\sigma$ confidence level for the thermal inertia.
 The VISIR data are better matched by a relatively high roughness.
 The derived corresponding values are then D$_{eff}$ = 508$^{+3}_{-21}$\,m
 (at $\rho$ = 1.0, $\Gamma_{optimum}$ = 850\,Jm$^{-2}$s$^{-0.5}$K$^{-1}$).
 The equivalent size of 1999~RQ$_{36}$ (of an equal volume sphere)
 is therefore in the range between 480 to 511\,m.
 The true size error is probably a bit larger because
 the equator-on observing geometry does not allow us to ``see" the pole regions
 in the mid-to-far-IR thermal emission.

 The derived diameter is significantly smaller than the 610\,m
 derived from Spitzer measurements alone (Emery et al.\ \cite{emery10}).
 Such a large size is clearly not compatible with the full set of 
 thermal measurements. Emery (private communication) confirms that
 the published size was produced by a simple thermal model (NEATM,
 Harris \cite{harris98}), which is known to overestimate the diameters
 of asteroids by 15-20\% especially when the thermal dataset is
 limited to only large phase angles (Harris \cite{harris06}).

 It is interesting to note that Hudson et
 al.\ (\cite{hudson00}) used the first Goldstone radar images
 -taken in 1999- to derive a diameter estimate of $\sim$500\,m, very close to
 our value. Their preliminary results indicate a remarkably featureless
 surface (down to the resolution limit of 19\,m) and an almost
 spherical object, but they derived a wrong spin period of $\sim$2\,hr.
 Later on, Nolan et al. (\cite{nolan07}) used multiple radar data from
 1999 and 2005, some of them taken with unusually good aspect coverage,
 and conclude from the combined dataset that 1999~RQ$_{36}$
 must be an irregular spheroid with about 580\,m in diameter, more than
 15\% larger than our value. Only the recent work by Nolan et al.\
 (\cite{nolan12}) has brought the size back to the 500\,m value.

\subsection{Geometric albedo}

 The derived albedo 
 is in the mid-range of values suggested for B-type asteroids
 (Tholen \& Barucci \cite{tholen89}). But, as noted before, the albedo value
 depends directly on the H-magnitude. Our solution of 0.045$^{+0.015}_{-0.012}$
 already includes the error of $\pm$0.3 in H-magnitude. If
 it turns out that the opposition effect of 1999~RQ$_{36}$ is steeper
 or shallower than assumed for calculating of the H-magnitude,
 then the derived geometric albedo would also change.
 Hergenrother et al.\ (\cite{hergenrother12}) provide, in addition to the above
 H-magnitude, H$_V$ and G by strictly following
 the IAU H-G photometric system for airless bodies
 (Bowell et al.\ \cite{bowell89}) and find H$_V$=19.90$\pm$0.10\,mag and
 G=-0.14$\pm$0.02. The corresponding radiometric size would not change, but
 the geometric albedo could then be as high as p$_V$=0.085.
 The low albedos for near-Earth objects are believed to be connected to
 a primitive volatile-rich surface
 composition (Fernandez et al.\ \cite{fernandez05}). But a dedicated
 orbital evolution and thermal history study by Delbo \& Michel (\cite{delbo11})
 has shown that 1999~RQ$_{36}$ has approached the Sun 
 in the past to regions of q $\le$ 0.8 (corresponding to about 400\,K at the
 sub-solar point) with an 80\% probability and with a lower
 probability even to q $\le$ 0.6 (480\,K at the sub-solar point). Such high
 temperatures alter the pristine properties of the surface material, because
 some of the expected primitive compounds already break up at moderate
 temperatures above 300\,K, and the surface volatiles might be completely lost
 (e.g., Marchi et al.\ \cite{marchi09}). But the temperature maxima at 3-5\,cm
 below the surface are already lower by 100\,K, and organics have very likely
 been protected from thermal break-up (Delbo \& Michel \cite{delbo11}).
 
\subsection{TPM concept}

 The calculations for the TPM beaming function
 for a rough surface are, strictly speaking, only applicable
 to the sunlit part of the asteroid (Lagerros \cite{lagerros98}).
 A change in roughness changes the radiation
 field. A rougher surface has the tendency to emit or ``beam" more
 in the solar direction at the expense of emission into larger
 phase angles. And this calculation is only done for the sunlit
 part of the surface.
 But even at large phase angles, the observed mid- and far-IR thermal
 emission is still dominated by the emission from the sunlit part
 of the object ($\sim$T$^4$), the emission from the dark and cold part beyond
 the terminator is negligible with or without roughness.
 Only for sub-milli\-metre/milli\-metre/centi\-metre wavelengths observations
 at very large phase angles above 90$^{\circ}$ are the night side contributions
 not negligible anymore, and these assumptions might turn out to be
 problematic and might result in wrong predictions.
 M\"uller et al.\ (\cite{mueller05}) used observations
 up to 110$^{\circ}$ phase angle to derive highly accurate size and
 albedo values for \object{25143~Itokawa}. The corresponding roughness
 and thermal inertia settings (similar to the values found here)
 explain the observations taken at 27$^{\circ}$ and 54$^{\circ}$
 with a similar accuracy to the observations taken at 108-110$^{\circ}$,
 giving us confidence that our TPM setup is appropriate also for
 the 1999~RQ$_{36}$ observations.

\section{Conclusions}
\label{sec:conclusion}
 
 The Herschel DDT measurements of 1999~RQ$_{36}$,
 combined with ESO-VISIR DDT measurements and data from the
 Spitzer Space, have allowed us to conclude that this asteroid has a
 diameter of an equal volume sphere in the range 480 to 511\,m, considerably smaller than
 previously estimated radar and radiometric values (Emery et al.\ \cite{emery10}; 
 Nolan et al.\ \cite{nolan07}), but in good agreement with the recent mean diameter
 given by Nolan et al.\ (\cite{nolan12}) of 493$\pm$20\,m. The thermal
 lightcurve variations are best-fit by an ellipsoidal shape model with a
 best fit a/b-ratio of of 1.04, corresponding to an ellipsoidal 
 body with 509 $\times$ 489 $\times$ 489\,m ($\pm$ about 10\,m in $a$- and $b$-
 and an unknown error in $c$-dimension).
 1999~RQ$_{36}$ has a geometric albedo p$_V$ = 0.045$^{+0.015}_{-0.012}$,
 in the nominal range for primitive volatile-rich NEAs of this type,
 and has a thermal inertia $\Gamma$=650 $\pm$ 300\,Jm$^{-2}$s$^{-0.5}$K$^{-1}$,
 similar to Itokawa, a rubble pile asteroid.
 Our results further serve to confirm that the asteroid has a
 retrograde rotation with a spin vector between -70 to -90$^{\circ}$.
 The inconsistency between the flux value derived from VISIR observations
 and the Spitzer-IRS dataset, taken at very different observing geometries,
 suggests there may be large variations in roughness on the surface.
 If the motion of surface material or regolith turnover processes are responsible
 for different terrains, then one can expect to find regions
 on the surface with very fresh material, at least at subsurface layers
 at several centimetres depth.
 The results of this work on the hypothesis that two different types of 
 terrains exist with different levels of roughness and the possibility
 of having part of the surface with fine dust and pebbles is very
 important for the OSIRIS-REx mission sample mechanism strategy.
 A global mapping during the
 presampling phase will then allow the OSIRIS-REx team to select the
 most optimal area for collecting pristine and organic-rich material.

 With this project we explore possibilities for thermophysical
 model techniques for an individual target that will soon be
 seen and characterised in detail. It will allow us to verify
 our findings on physical and thermal properties, as well as the
 proposed different surface terrains that are needed to explain
 the derived roughness and thermal inertia. Radiometric techniques
 are very powerful for deriving highly reliable size and albedo
 information, but signatures of surface texture, shape, spin-state,
 and thermal properties are also included in the object's thermal
 emission. In fact, because thermal data is (or will soon be) available
 for many thousands of asteroids (IRAS, MSX, ISO, Akari, Spitzer,
 WISE, ground-based mid-IR/submm/mm programmes, etc.), the results
 from 1999~RQ$_{36}$ can be considered to be the key to allowing
 us to transfer our model techniques to many other targets that
 will not be visited by spacecraft in the near future, but for which similar
 important questions can now be answered.

\begin{acknowledgements}
 HSpot \& HIPE are joint developments by the Herschel Science
 Ground Segment Consortium, consisting of ESA,the NASA Herschel
 Science Center, and the HIFI, PACS, and SPIRE consortia.
 We gratefully acknowledge the awarding of director discretionary
 time from the Herschel Space Observatory and ESO to support the work published here.
 We would also like to acknowledge the generous support by Mario van den Ancker (ESO)
 for installating and running of the VISIR pipeline and
 the support by Eric Pantin (CEA) and Ralf Siebenmorgen (ESO)
 for VISIR calibration and error calculation issues.
 Support for the Spitzer data reduction was provided by
 Sonia Fornasier, Josh Emery, and Jeffrey Van Cleve.
 It is also our pleasure to thank the Herschel Science Centre
 Mission Planning Team for their efforts in the
 optimal scheduling of these observations. Cs.\ K.\ and A.\ P.\ are supported
 by the ESA Grant PECS-98073 and by the Bolyai Research Fellowship
 of the Hungarian Acedemy of Sciences. We would also like to thank
 the anonymous referee for a very thorough and very
 constructive review of the manuscript.
\end{acknowledgements}

\end{document}